\documentstyle[preprint,pre,aps]{revtex}
\tightenlines
\begin{document}

\title{Time decay of the remanent magnetization\\
in the $\pm J$ spin glass model at $T=0$ }
\author{ J. K\l{}os$^{1,2}$ and S. Kobe$^{1}$}
\address{$^{1}$Institut f\"ur Theoretische Physik,
Technische Universit\"at Dresden,\\ D-01062 Dresden, Germany\\
$^{2}$Institute of Molecular  Physics, Polish  Academy  of  Sciences\\
Smoluchowskiego  17/19, 60-179 Pozna\'n, Poland\\
 }
\date{\today}
\maketitle
\begin{abstract}

Using the zero-temperature Metropolis dynamics, the time decay of the remanent magnetization in 
the $\pm J$ Edward-Anderson spin glass model with a uniform random distribution of ferromagnetic and 
antiferromagnetic interactions has been investigated. Starting from the saturation, the magnetization per spin 
$m$
reveals a slow decrease with time, which can be approximated by a power law:
$m(t)= m_{\infty} + \left({t\over a_{0}}\right)^{a_{1}}$, $a_{1} < 0$.
Moreover, its
relaxation does not lead it into one of the ground states, and therefore the system is trapped
in metastable isoenergetic microstates remaining magnetized. Such behavior 
is discussed in terms of a random walk the system performs on its available configuration space. 
\end{abstract}
\vskip1pc
\draft
%{\it Keywords:} Spin glass; Frustration; Remanent magnetization; Computer
%simulation} \vskip2pc
%%%%%%%%%%%%%%%%%%%%%%%%%%%%%%%%%%%%%%%%%%%%%%%%%%%%%%%%%%%%%%%%%%%%%
\section{INTRODUCTION}
\label{leb:Introduction} 

Spin glasses are systems that, at temperatures below so-called glass transition temperature $T_{g}$,
 find themselves in states with frozen disorder, i. e.,
 no long-range patternlike order typical for ordered magnets is present. 
Their properties are determined by competing ferromagnetic and antiferromagnetic exchange
interactions that are randomly distributed in the system.
Both the competition among the different interactions between the magnetic
moments and their random distribution all over a given system are likely to contribute significantly to such
unusual glassy behavior. 
From the theoretical point of view, they can be discussed in terms of the coarse-grained
free energy. Namely,  due to accidental degeneracy present in such systems, below $T_{g}$ their free 
energy landscape becomes extremely rough, with many local minima
corresponding to the same macroscopically observed properties but with entirely different microscopic states in the
system phase space. The minima (valleys) are separated from each other by some energy barriers, 
and once a system finds itself in one of them, it might take a lot of time on 
laboratory time scales to transit to the others. Thus, the observed  
properties of spin glasses may only correspond to those of one single valley in which the 
system happens to 
be, and as a result, ergodicity is practically broken~\cite{BinderYoung,Fischer}.
In other words, spin glasses can be seen as systems whose dynamics at low temperatures is extremally slow and
whose properties measured in real experiments always correspond to situations out of equilibrium.        
Very good evidence for this nonstationary dynamics is 
the response of the system ac susceptibility to an oscillating field, i.e., its dependence both on time
and frequency.
Another example is the slow decay of remanent
magnetizations with time~\cite{BinderYoung,Fischer}. The thermoremanent magnetization 
(TRM) is measured by cooling the sample in a nonzero magnetic field $H$ from above the glass-transition
temperature $T_{g }$ to a temperature $T$ below it and then swiching off the field. The isothermal remanent
magnetization (IRM) is measured by zero-field cooling of the sample in the same way as before, then turning on 
the field, and subsequently turning it off.  In addition, in both cases the experimental results show that
for small fields applied to the sample, the remanent magnetization is strongly affected by the so-called 
waiting time at which the sample is kept at constant temperature before the external field 
is changed~\cite{Norblad}.
According to the decay of
both the remanent magnetization and the energy, the experimental results depend on the observation 
times after the field is switched off in all measurements.
The above nonstationary dynamics has been described by a fair variety of functions. 
The most important ones
include power law, logarithmic, stretched exponential, and others, and 
the question of judging which is the most universal is still far from decided~\cite{Norblad,Ferre,Chamberlin}.
Additionally, as some experiments
indicate, the remanent magnetization decays so slowly with time that some nonzero remanence is still observed
over macroscopic time scales, particularly at very low temperatures~\cite{Blinc}. Relaxation time
measurments in CeNi\raisebox{-.6ex}{\scriptsize 0.8}Cu\raisebox{-.6ex}{\scriptsize 0.2} below the spin glass temperature 
6 K show that the decay time
increases drastically with a distinct tendency to a state with nonzero magnetization, which is
higher the lower the temperature is~\cite{Blanco}. The theoretical background of such a property 
is rather unclear~\cite{Blinc}.
  
One of the simplest theoretical models of spin glasses is the Edwards-Anderson (EA) model. Ising spins 
are located at each site of a lattice with randomly distributed
ferromagnetic and antiferromagnetic interactions between the nearest neighbors. 
Such a model
reveals most of the crucial features typical of real spin glasses including relaxation
phenomena.  
Both the early papers~\cite{Binder,Kinzel1} and the newer ones~\cite{Kinzel2,ParisiRitort}
on two- (2D) and three-dimensional (3D) models with a Gaussian distribution of bonds confirm 
that in a wide range of temperatures, a remanent magnetization occurs
that decays very slowly with time $t$ according to
a power law $m(t)\sim {t^{-\alpha}}$ with $\alpha(T) \sim T$.
It it also known that in case the couplings among spins may take only discrete 
values ($\pm J$ models) 
at sufficiently low temperatures, relaxation properties of such glasses are entirely 
different
because of the existence of energy gaps in their energy spectra~\cite{ParisiRitort,Kirk1}.
As a result, for models with bimodal distribution of interactions at temperatures well below $T_g$, a simple 
linear dependence of $\alpha$ on temperature $T$ is not satisfied, and the functional form of the remanent 
magnetization decay is still rather far from being established. It has been mentioned that
the function $\alpha(T)$ might go to zero faster than linear at low temperatures~\cite{ParisiRitort}.
It has also been suggested that at finite
small temperatures, $m(t)$ should coincide with $m(t)$ at zero temperature for some
region of time, and that the relaxation causes such a system to remain trapped in one
metastable state with a finite remanent magnetization.  

In this paper, we revisit the remanent magnetization decay of 2D $\pm J$ EA spin glass 
because we are interested in its extremely low-temperature relaxation properties where 
the power law with $\alpha(T) \sim T$ breaks down.
We consider the limiting case of zero-temperature behavior in order to find out whether 
the remanence phenomena observed at such conditions could reflect the low-temperature 
properties of discrete systems at least qualitatively. We would also like to check out 
whether or not results obtained at $T=0$ could be treated as continous extrapolations of those at low but 
finite temperatures, which, according to similar research done on SK model~\cite{Eissfeller}, seems rather unlikely. 
Actually, in~\cite{Eissfeller} it is shown that even the zero-temperature dynamics provides a 
decay of the magnetization that can be fitted by a power law with a constant
exponent $\alpha$.
We carry out simulations using
the zero-temperature dynamics, which has been successfully applied to various
spin lattices including the persistence probability in weakly disordered Ising model~\cite{Jain}, 
hysteresis in the random-field Ising model on the Bethe lattice~\cite{Sethna}, the question of avalanches in 
spin systems~\cite{Perkovic}, and others~\cite{Acharyya}.     
We investigate the
system relaxation process towards low-energy states by plotting both energy and magnetization versus the
zero-temperature Monte Carlo steps (MCS) per spin, which are treated as "time units" $t$. 
We discuss the influence of frustration on the nonequilibrium time properties 
by comparing them with corresponding properties for an unfrustrated system with pure antiferromagnetic 
interactions. 
Finally, we discuss the obtained results in terms of the system random walk on its configuration space.          

%%%%%%%%%%%%%%%%%%%%%%%%%%%%%%%%%%%%%%%%%%%%%%%%%%%%%%%%%%%%%%%%%%%%%%%%%%%%%%%%%%%
\section{MODEL AND SIMULATION}
\label{leb:Description} 

We use the Edward-Anderson spin glass model with a random and uniform distribution
of discrete interactions $J_{ij}=\pm 1$ between the nearest neighbors all over a 2D square lattice with $N$ sites and
with periodic boundary conditions. 
The Hamilton function of such a system in an external magnetic field $B$ is:
\begin{equation}
H = -\sum _{i < j}J_{ij}S_{i}S_{j}-B\sum _{i}S_{i}, 
\label{eq:hamiltonian}
\end{equation}
where $S_{i}, S_{j}=\pm 1$ (up/down) are Ising spins, and the sum in the first term on the right-hand side runs over
the nearest-neighboring lattice sites. The samples are prepared in such a way that the fraction of 
antiferromagnetic bonds is $p = 0.5$. We study how the remanent magnetization decays
with time in zero magnetic field starting from the system saturation state. The magnetization per 
spin is given by
\begin{equation}
m(t) = (N_{+}(t)-N_{-}(t))/N,
\label{eq:magnetization}
\end{equation}  
where $N_{+}$, $N_{-}$ denote the number of spins up and down, respectively. 
The system relaxation process is simulated by
applying a version of the zero-temperature Metropolis algorithm given by the 
following steps, cf.~\cite{Sethna,Acharyya}:\\   
(i) Consider a sample in its saturation state (all spins up).\\ 
(ii) Pick a spin at random.\\
(iii) Flip it only if this process does not increase the energy.\\
(iv) Repeat steps (ii) and (iii) $N$ times. So, one time unit is defined as 
   one MC step per spin.\\
(v) Record magnetization and energy as functions of these time units.\\ 
(vi) Start again with step (ii) to find results for the following time step.\\
Thus, the above algorithm determines the system
random walk on the configuration space in the direction of dropping energy. It should be mentioned that a different
algorithm based on the Glauber dynamics is also used in the literature, cf.~\cite{Jain,Newman,Newman1}. Then for $T=0$ 
step (iii)
of the above mentioned procedure is replaced by the following one: Flip it if this process 
decreases the energy and flip it with the
probability $1/2$ if this process does not change the energy.
%%%%%%%%%%%%%%%%%%%%%%%%%%%%%%%%%%%%%%%%%%%%%%%%%%%%%%%%%%%%%%%%%%%%%%%%%%
\section{RESULTS}
\label{leb:Results}  

First we consider a special system of the size $N=70\times70$, for which the exact ground-state energy
is calculated using a branch-and-cut algorithm by De Simone {\it et al.}~\cite{DeSimone}. Its energy and 
magnetization decays 
are compared with those of an unfrustrated one of the same size with solely antiferromagnetic interactions
between the nearest neighbors.  
No qualitative differences in the energy relaxation can be seen, except for the fact that the ground-state energies
of both systems are different (Fig. 1).  
On the other hand, with respect to the remanent magnetization decay, both
systems are entirely different and the influence
of randomness and frustration on the phenomena can be observed. While the unfrustrated system
moves extremely fast to the low-energy unmagnetized states, the frustrated one relaxes more slowly
and the isoenergetic states it finally explores still possess nonzero magnetization (Fig. 2). 
In that area, the remanent magnetization fluctuates around a constant value, which refers to simply flipping idle
spins.

In order to describe the system dynamics more quantitatively, we have considered $10$ samples of the size
$N=50\times 50$ with different distribution of bonds. Typically, we have performed around 2000 MC relaxation steps.  
In Fig. 3, we show the remanent magnetization $m(t)$ averaged over the samples and over $30$ independent runs for each. 
The best fit for the dependence of the magnetization on time (for $t>0$) is obtained by
\begin{equation}
m(t)= m_{\infty} + \left({t\over a_{0}}\right)^{a_{1}}, a_{1} < 0.
\label{eq:algebraicfit}
\end{equation}
The first term $m_{\infty}$ corresponds to
the average remanent magnetization of the system trapped in a subspace of low-lying isoenergetic states after 
a long time of decay, whereas the second term corresponds to the nonstationary dynamics. 
The parameters found are $m_{\infty}=0.059$, $a_{0}=0.23$, $a_{1}=-0.83$. The insert of Fig. 3 shows 
$\ln[m(t)-m_{\infty}]$ versus $\ln(t)$ with $m_{\infty}$ taken from the power-law fit (\ref{eq:algebraicfit}). The 
observed dispersion of the computed points for 
large values of time $t$ is due to their small fluctuations around $m_{\infty}$. 
We have also tested some other functions with four 
free parameters to adjust the data. Among them, the stretched exponential law
$m\left( t \right) = m_{\infty}+a_{0}\exp\left[-\left(t/a_{1}\right)^{a_{2}}\right]$ (for $t\ge0$), 
with $m_{\infty}=0.060$, $a_{0}=0.94$, $a_{1}=0.43$, and $a_{2}=0.35$, has proved reasonably close to it, however a 
bit worse than the power-law one.

\section{Summary}
\label{leb:Summary}  
We have investigated the remanent magnetization decay of a 2D $\pm J$ EA spin glass model at zero temperature.
All the samples were initially at their saturation states and then allowed to relax towards states with 
lower energy using the zero-temperature Metropolis algorithm. 
The observed remanent magnetization decay in the frustrated system was much slower than in the 
unfrustrated one, whose remanent magnetization decreased rapidly to zero.
After a number of MC steps, the walk of the random system on its configuration space 
practically became limited to subspaces of magnetized isoenergetic states. At that region, on the average, the 
remanent magnetization fluctuated without further decrease. 
Our calculations suggest that the time decay of the the remanent magnetization can be very well represented 
by a simple power-law formula with three fitting parameters. 
From a microscopic point of view, the nonexponential relaxation phenomena in spin glasses can be 
discussed in terms of random diffusion on the available configuration space.  
It is suggested by simulations that stretched exponential relaxation behavior in glassy
systems appears with the exponent $a_{2}$ going to $1/3$ when approaching a percolation transition
in the configuration space, which is  
a multidimensional 
hypercube~\cite{Campbell1,Campbell2,Campbell3}.
The same form of the magnetization decay in a $2D$ ferromagnetic Ising
model has been reported and some dependence of the 
relaxation phenomena on the system dimension has been found as well~\cite{Stauffer,Ogielski,Takano}.
In the so-called trap model, the system evolves among various 
traps with random 'trapping times'. The traps are separated from each other by energy barriers that can be 
crossed by thermal excitations~\cite{Bouchaud1,Bouchaud2}. 
This kind of approach has been successfully used to study low-temperature aging of a 
system consisting of configurations with random energies~\cite{Barrat}.  
However, since the dynamics of ours is athermal, there is no such barrier crossing in it and 
a purely entropic interpretation might be adapted instead~\cite{Bouchaud2,Franz}. 
Although for simulations at finite temperatures a mixture of 'energetic' and 'entropic' barriers is likely
to contribute to the phenomena of slow dynamics, one of 
the effects vanishes with respect to the zero temperature simulations.
The observed slowing down 
of the system relaxation could be qualitatively understood with the help of so-called entropic traps themselves. 
While the point in the
configuration space randomly searches for available paths leading to states with lower energy, 
their number decreases rapidly with time. As a result, we find that the lower the energy of the states is that 
the system has reached the longer is the time period needed to leave them. The phenomena 
is also very consistent with the Markov theory~\cite{Feller,Kubik}.
It says that every finite Markov chain contains at least one absorbing set in which the system remains forever after
falling in it (a good example of an absorbing set is the closure of one of the ground states). Moreover, the
theory states that the probability of passing to one of them with 'time' going to infinity tends to one. 
That is, after a large number of MC steps, the available configuration space of the system becomes one of the
absorbing sets, where the energy is kept constant and the  magnetization fluctuates. A finite value of remanent
magnetization $m_{\infty}$ indicates that the absorbing set contains excited states rather than ground states.
This is consistent
with recent results on magnetic hysteresis at zero temperature~\cite{Vogel}. All metastable states connected
by one-spin flips without raising energy form a local ensemble of metastable states. When all these states
belong to a relatively large value of $m$, the one-spin interconnection with the local ensemble of
ground states can be prohibited.

Moreover, our simulations confirm that the observed tendency of the exponent $\alpha(T)$ 
[in the formula $m(t)\sim {t^{-\alpha}}$] to go to zero with dropping temperature seems to be in 
direct contradiction to the remanent magnetization decay at exactly $T=0$, for which $\alpha$ is finite.
This also seems to indicate that the violations of the power law at very low temperatures might be due
to the characteristic time $\tau$ divergence~\cite{ParisiRitort} rather than to the function $\alpha(T)$ going
to zero faster than linear. Moreover, we think that another explanation 
could be suggested as well. Namely, it has been experimentally found that the remanent magnetization $m_{\infty}$     
decreases with increasing temperature~\cite{Blinc,Blanco}. Below a certain value of temperature $T$, the discrete
structure of the system energy spectrum is likely to affect its properties significantly and $m_{\infty}$ may 
become relevant. That means that at very low temperatures it should be contained in the 
relaxation law, and a plot $\ln[m(t)-m_{\infty}]$ versus $\ln(t)$ (instead of $\ln[m(t)]$ versus $\ln(t)$)
could still remain linear.  

\acknowledgments
Support from Graduiertenkolleg "Struktur- und Korrelationseffekte in Festk\"orpern" is gratefully acknowledged.
We thank D. Stauffer, J. Wei{\ss}barth, and M. Nogala for helpful discussions, and the
authors of Ref.~\cite{DeSimone} for allowance to use their computer data.
%%%%%%%%%%%%%%%%%%%%%%%%%%%%%%%%%%%%%%%%%%%%%%%%%%%%%%%%%%%%%%%%%%%%
  
\begin{figure}
\vspace{1cm}
\caption{Energy per spin $E$ (in units of $|J_{ij}|$) versus time $t$ 
(in units of Monte Carlo steps per spin) averaged over $30$ independent runs in
a frustrated $N=70\times70$ system (circles) and in an unfrustrated one of the same size with solely
antiferromagnetic interactions (triangles). The dotted horizontal 
line represents the exact ground-state energy of the former. Energy axis is shifted by adding a term 2.}
\label{fig:Energies70x70}
\end{figure}
\begin{figure}
\vspace{1cm}
\caption{Magnetization per spin $m$ versus time $t$ (in units of Monte Carlo steps per spin) averaged 
over $30$ independent runs for the same systems as in Fig. 1.}
\label{fig:Magnetizations70x70}
\end{figure}
\begin{figure}
\vspace{1cm}
\caption{Magnetization per spin $m$ versus time $t$ (in units of Monte Carlo steps per spin)  
averaged over $10$ samples of the size $N=50\times50$ and over $30$ runs for each. The continous line 
represents the fit according to the power law (\ref{eq:algebraicfit}). The inset shows $\ln[m(t)-m_{\infty}]$ versus
$\ln(t)$.}
\label{fig:Bestfit}
\end{figure}
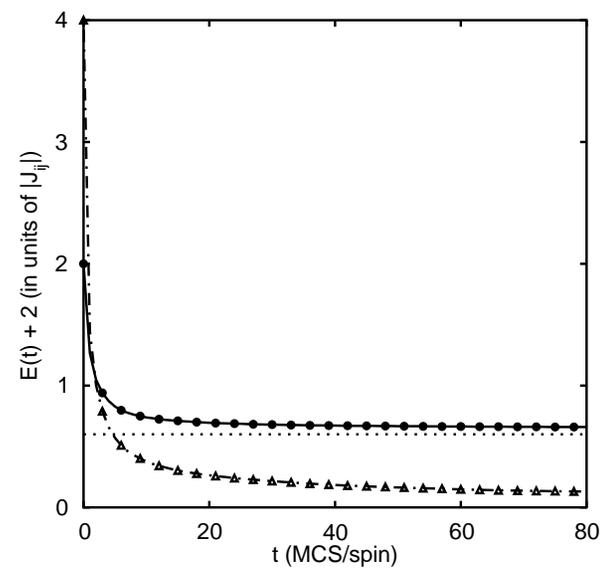
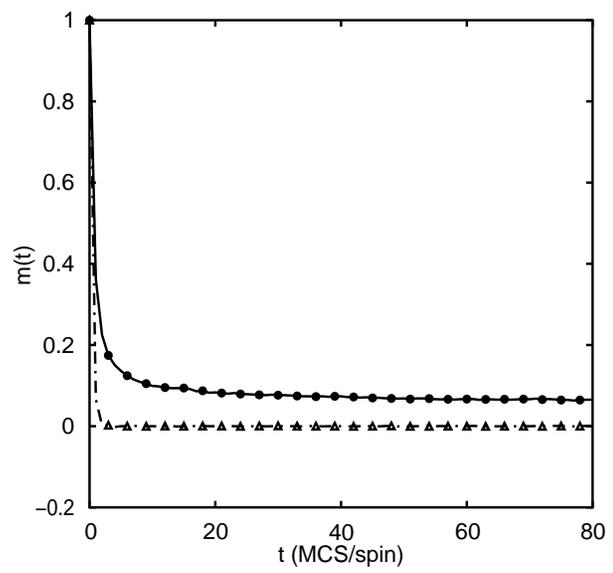
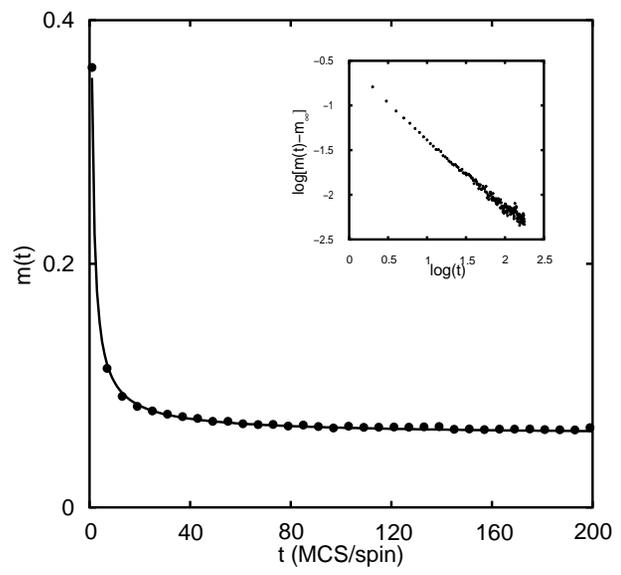
\end{document}